\pdfoutput=1

\documentclass[showpacs,nofootinbib,showkeys,eqsecnum,prd,aps,preprint]{revtex4-2}

\usepackage[english]{babel}

\usepackage{amssymb}
\usepackage{amsmath}
\usepackage{amsfonts}
\usepackage{latexsym}
\usepackage{mathrsfs}
\usepackage{bm}
\usepackage{tensor}
\usepackage{hyperref}
\usepackage{graphicx}

\def\Sp{\mathop{\rm Sp}\nolimits\,}
\def\const{\mathop{\rm const}\nolimits\,}
\def\dac{\displaystyle\frac}
\def\dil{\displaystyle\int\limits}
\newtheorem{theorem}{Theorem}

\begin{document}

\title{Semiclassical approach to the nonlocal kinetic model of metal vapor active media}

\author{Alexander V. Shapovalov }
\email{shpv@phys.tsu.ru}
\affiliation{Department of Theoretical Physics, Tomsk State University, Novosobornaya Sq. 1, 634050 Tomsk, Russia}
\affiliation{Laboratory for Theoretical Cosmology, International Centre of Gravity and Cosmos, Tomsk State University of Control Systems and Radioelectronics, 40 Lenina av., 634050 Tomsk, Russia}

\author{Anton E. Kulagin}
\email{aek8@tpu.ru}
\affiliation{Laboratory of Quantum Electronics, V.E. Zuev Institute of Atmospheric Optics, SB RAS, 1 Academician Zuev Sq., 634055 Tomsk, Russia}
\affiliation{Division for Electronic Engineering, Tomsk Polytechnic University, 30 Lenina av., 634050 Tomsk, Russia}

\begin{abstract}
A semiclassical approach based on the WKB-Maslov method is developed for the kinetic ionization equation in dense plasma with approximations characteristic of metal vapor active media excited by a contracted discharge. We develop the technique for constructing the leading term of the semiclassical asymptotics of the Cauchy problem solution for the kinetic equation under the supposition of weak diffusion. In terms of the approach developed, the local cubic nonlinear term in the original kinetic equation is considered in a nonlocal form. This allows one to transform the nonlinear nonlocal kinetic equation to an associated linear partial differential equation with a given accuracy of the asymptotic parameter using the dynamical system of moments of the desired solution of the equation. The Cauchy problem solution for the nonlinear nonlocal kinetic equation can be obtained from the solution of the associated linear partial differential equation and some algebraic equations for the coefficients of the linear equation. Within the developed approach, the plasma relaxation in metal vapor active media is studied with asymptotic solutions expressed in terms of higher transcendental functions. The qualitative analysis of such the solutions is given.\\
\end{abstract}



\maketitle

\section{Introduction}
\label{sec:int}

 Studies of kinetics of metal vapor active media (MVAM) are motivated by their wide application in the development of laser systems. MVAM are used in technics due to their high optical gain in a narrow spectral range \cite{kazaryan02,asratyan16,klyuch19}. Nowadays, the most promising application of MVAM are the active optical systems (the so-called laser monitors) that allow one to visualize the processes blocked by the intense strong background light \cite{lasmon14,lasmon16}.

 \par

 The active media on metal vapors are a mixture of a buffer gas (inert gas) and a gaseous metal, and the concentration of the buffer gas is 2-3 orders of magnitude higher than the concentration of metal vapors. Under the action of an electric discharge, the processes of ionization and recombination in such media occur mainly due to the reactions of electron impact. The inelastic collisions of neutral atoms with electrons are responsible for ionization, and triple recombination processes (triple collision of an ion with two electrons) are responsible for deionization. In MVAM, mainly metal atoms are ionized, and the buffer gas practically does not contribute to the concentration of the electron gas due to the much higher ionization energy. In a number of works (see, for example, \cite{kulopt17,gubarev16,behrouzina19,boichenko05,marshall04}) active media were investigated, where neon acted as a buffer gas,
and vapors of copper and its halides did as an active substance.

Note that a mixture of a buffer gas and metal vapors in this case
is inside the gas discharge tube (GDT). However, under the condition
of a strongly contracted pumping discharge of the active medium, the
ions will be localized around the center of the GDT and there will
be no boundary conditions on the GDT walls in the mathematical problem
statement. Here we will consider just such a case when the ion concentration
rapidly decreases with distance from the center of the GDT. The equation
for the concentration of positive singly charged ions $n_{i}(\vec{x},t)$
of a metal for a constant gas temperature according to the law of
mass action can be written as \cite{kushner83}
\begin{equation}
\begin{gathered}
\partial_t n_i(\vec{x},t)=D_a(t)\Delta  n_i(\vec{x},t) + q_i(\vec{x},t) n_e(\vec{x},t) n_{neut}(\vec{x},t) - \\
-q_{tr}(\vec{x},t)n_i(\vec{x},t)\big(n_e(\vec{x},t)\big)^2,
\end{gathered}
\label{fkpp1}
\end{equation}
where the space and time variables are denoted by $\vec{x\,}(\in{\mathbb{R}}^{n})$
and $t\,(\in R^{1})$, respectively, $\partial_t=\partial/\partial t$; $q_{i}(\vec{x},t)$ is rate constant
of the electron impact ionization process, and $q_{tr}(\vec{x},t)$
is the rate constant of the triple recombination process. The ambipolar
diffusion coefficient is $D_{a}(t)$; the concentration of neutral
metal atoms is $n_{neut}(\vec{x},t)$, and $n_{e}(\vec{x},t)$ is
the concentration of electrons. The dependence of the coefficients
$q_{i}$, $q_{tr}$ and $D_{a}$ on $\vec{x}$ and $t$ is due to
their dependence on the electron temperature which can be substantially
inhomogeneous in time and space. When the discharge energy is insufficient
for the complete ionization of metal vapors, then only singly charged
ions are produced in the plasma. Therefore, in view of the plasma
quasineutrality, the concentration distribution of positive singly
charged ions, $n_{i}(\vec{x},t)$, coincides with the concentration
distribution of electrons, $n_{e}(\vec{x},t)$, i.e.,
\begin{align}
n_{e}(\vec{x},t)=n_{i}(\vec{x},t).
\label{fkpp1a}
\end{align}
The properties of active media that are useful for applications appear when the upper resonance energy level of metal atoms is effectively pumped. In such the conditions, the degree of ionization is small, i.e.
\begin{align}
n_{neut}(\vec{x},t)\gg n_{i}(\vec{x},t),
\label{fkpp1b}
\end{align}
and $n_{neut}$ almost does not depend on $n_i$. In practice, $n_{neut}$ is at least one order greater than $n_i$ for MVAM. Under conditions \eqref{fkpp1a}, \eqref{fkpp1b}, the equation \eqref{fkpp1} becomes closed and can be written as
\begin{equation}
\begin{gathered}
\partial_t n_i(\vec{x},t)=D_a(t)\Delta  n_i(\vec{x},t) + a(\vec{x},t) n_i(\vec{x},t) -q_{tr}(\vec{x},t)n_i^3(\vec{x},t),
\end{gathered}
\label{fkpp1c}
\end{equation}
where $a(\vec{x},t)=q_i(\vec{x},t) n_{neut}(\vec{x},t)$, $q_{tr}(\vec{x},t)$, and $D_a(t)$ are given functions. For $D_a=\const$, $a=\const$, $q_{tr}=\const$, the equation \eqref{fkpp1a} is termed the Newell--Whitehead equation \cite{newell69,vaneeva19}. The kinetic equation with a cubic nonlinearity of the form {\eqref{fkpp1c}} have applications going beyond the plasma physics. For example, it can be treated as dissipative part of the Gross--Pitaevskii equation with a phenomenological damping that describes the formation of vortices in Bose--Einstein condensates \cite{Ji2008,physrev1} or as a model equation for the imaginary-time method of constructing stationary solutions of the Gross--Pitaevskii equation \cite{Wang2010,amara93,Liang2005}.\\

We assume in \eqref{fkpp1} that diffusion and ionization/recombination
processes occur at different scales in spatial coordinates. This approximation
is applied when the electron temperature has a weak spatial inhomogeneity.
The ambipolar diffusion coefficient $D_{a}(t)$ is assumed to be $D_{a}(t)=D_{i}\Big(1+{\displaystyle \frac{T_{e}(t)}{T_{g}}\Big)}$
where $T_{e}(t)$ is the electron temperature, $T_{g}$ is the gas
temperature, and the ion diffusion coefficient $D_{i}$ is independent
of the spatial variables. The dependence of $q_{i}$ and $q_{tr}$
on the electron temperature is stronger than that of $D_{a}$. Therefore
even a weak dependence of the electron temperature on the spatial
variables can lead to a significant dependence of $q_{i}$ and $q_{tr}$
on $\vec{x}$. We do not take into account the dielectronic recombination
process in the equation \eqref{fkpp1} since it makes a significant
contribution to the ion concentration only in a rarefied plasma with
pressures much lower than those that are characteristic of the operation
of MVAM.   Also, we do not take into consideration the Penning ionization that is significant in MVAM where the buffer gas pressure under normal conditions is higher than one hundred torr while it is only 20-30 torr in most present-day works.

The coefficients $q_{i}(\vec{x},t)$ and $q_{tr}(\vec{x},t)$ in equation
(\ref{fkpp1}) mean the total rates of the corresponding processes
including stepwise ionization and also recombination to the lower
energy states of neutral atoms. This approximation allows one not
to solve a system of a large number of equations where each equation
describes a population of an individual energy level of neutral atoms.
This approximation is widely used in describing the ionization in
plasma, since direct experimental data usually give the values of
the total ionization rates (see, e.g., \cite{freund90}. The total rate of triple recombination is determined
on the basis of the semiclassical approach described in the work of
Gurevich and Pitaevskii \cite{gurpit64}.

Equation (\ref{fkpp1}) plays an important role in the design of MVAM. The influence of the prepulse electron concentration
on characteristics of active media on copper vapors was discussed
in detail in \cite{carman98,boichenko2001}. In
particular, solutions to the equation (\ref{fkpp1}) were required
to construct a high-voltage high-frequency pumping circuit for exciting
the active medium for laser monitors. The plasma itself has an active-inductive
resistance character. In this case, the active component of the resistance
prevails. This parameter significantly depends on time within the
pump pulse period and it is a complex function of the temperature
and electron concentration. Therefore, to match the pumping circuit
with the load, models of the resistance of the active medium are used.
These models include the electron concentration or, at least, its
prepulse value (see, e.g., \cite{kyureg19}).
The main way of determining it is related to solutions of kinetic
equations. The kinetic modelling of such active media began to develop in the 80-90s (see, e.g., \cite{kushner83,yurchenko84,carman94,cheng97}). The approach for constructing a space-time
kinetic model of active media on copper vapor was developed in \cite{kulopt20,kulopt19,kulphys18} where the model equations were studied mainly numerically.

The aim of this work is to develop an analytical approach based on
the WKB-Maslov theory of the semiclassical approximation \cite{Maslov1,Maslov2,BeD2}
to study the kinetic equation of plasma ionization.

The method of semiclassical asymptotics was applied in \cite{shap2009,LST14,fkppshap18} to
a nonlocal generalization of the Fisher-Kolmogorov-Petrovskii-Piskunov
equation known in the theory of biological populations, and also in
\cite{shapovalov:BTS1,sym2020,kulagin2021} for the nonlocal Gross-Pitaevsky equation which is widely
used in the theory of Bose-Einstein condensates. The approach proposed
here for the kinetic equation of plasma ionization essentially involves
the results of \cite{shap2009,LST14,fkppshap18}.

The paper is structured as follows. In Section \ref{sec:2}, we introduce basic notations and the problem setup. The class of semiclassically concentrated functions, where asymptotics are constructed, is presented. In Section \ref{sec:3ee}, the dynamical system describing the evolution of moments of the unknown solution is deduces and it is considered within the framework of our approach. In Section \ref{sec:linprob}, the family of associated linear equations is obtained. The leading term of an asymptotic solution to the original nonlinear kinetic equation is constructed from solutions of these equations according to the certain algebraic conditions. Section \ref{sec:example} illustrates general approach with the specific example of the plasma relaxation problem. In Section \ref{sec:concl}, the concluding remarks are given.

\section{Nonlocal kinetic equation and semiclassical approximation}
\label{sec:2}

To apply the method of semiclassical asymptotics in accordance with \cite{fkppshap18,sym2020}, we consider a nonlocal version of the kinetic equation \eqref{fkpp1}.

In the local equation \eqref{fkpp1}, triple recombination is described in terms of  a contact interaction model. If we introduce into the model the dependence of the probability of the act of triple recombination on the mutual arrangement of the particles participating in it, then we obtain a nonlocal generalization of the equation \eqref{fkpp1} of the form
\begin{equation}
\begin{gathered}
\partial_t n_i(\vec{x},t)=D_a(t)\Delta  n_i(\vec{x},t) + q_i(\vec{x},t) n_{neut}(\vec{x},t) n_e(\vec{x},t) - \\
-\varkappa\cdot n_i(\vec{x},t)\displaystyle\int\limits_{{\mathbb{R}}^n\times {\mathbb{R}}^n} b(\vec{x},\vec{y},\vec{z},t)n_e(\vec{y},t) n_e(\vec{z},t)d\vec{y}d\vec{z}.
\end{gathered}
\label{fkpp2}
\end{equation}
Here, the kernel $b(\vec{x}, \vec{y}, \vec{z}, t)$ of the integral term has the meaning of the probability density of the capture by an ion at the point $\vec{x}$ of an electron at the point $\vec{y}$ and an electron at the point $\vec{z}$ with subsequent triple recombination. For convenience, we have explicitly identified the normalization nonlinearity parameter  $\varkappa$.

The probability of triple recombination depends on the electron thermal velocity 
 and on the mutual distance between the electrons and the ion. Therefore, in specific examples of the equation \eqref{fkpp2}, we will assume
\begin{equation}
\begin{gathered}
b(\vec{x},\vec{y},\vec{z},t)=\tilde{b}(\vec{x},\vec{x}-\vec{y},\vec{x}-\vec{z},t),
\end{gathered}
\label{fkpp2a}
\end{equation}
where the dependence of $\tilde{b}(\vec{x},\vec{r}_1,\vec{r}_2,t)$ on $\vec{x}$ and $t$ is caused by its dependence on the electron temperature.

 Further,  we denote $D_a (t) = D \cdot\tilde{D}_a(t)$ in the equation \eqref{fkpp2}, where $D$ plays the role of an small diffusion parameter in the proposed method of semiclassical asymptotics, and the function $\tilde{D}_a(t)$ is considered given. In accordance with \eqref{fkpp1a}, \eqref{fkpp1b}, \eqref{fkpp1c},  we set $n_e=n_i$ in the equation \eqref{fkpp2}, denote $n_i(\vec{x}, t) = u (\vec{x}, t)$, and the equation \eqref{fkpp2} takes the form
  \begin{equation}
\begin{gathered}
\partial_t  u(\vec{x},t)=D \tilde{D}_a(t)\Delta  u(\vec{x},t) + a(\vec{x},t) u(\vec{x},t) - \\
-\varkappa u(\vec{x},t)\displaystyle\int\limits_{{\mathbb{R}}^n\times {\mathbb{R}}^n} b(\vec{x},\vec{y},\vec{z},t) u(\vec{y},t)u(\vec{z},t)d\vec{y}d\vec{z},\\
\end{gathered}
\label{fkpp3}
\end{equation}
where $a(\vec {x}, t)$, $b(\vec{x},\vec{y},\vec{z},t)$ are considered to be a given infinitely smooth functions with respect to spatial variables at each point $t$ that increase, as $|\vec{x}|\to \infty$, $|\vec{y}|\to \infty$, $|\vec{z}|\to \infty$, not faster than the polynomial.

We will seek solutions $u$ of the equation \eqref{fkpp3} in the class ${\mathcal P}_D^t$ of trajectory-concentrated functions
(TCFs) depending on the  parameter $D$ \cite{bagrov1,shap2009,LST14,fkppshap18}:
\begin{equation}
 {\mathcal P}_D^t=
\biggl\{\!\Phi :\Phi(\vec x,t,D)= \varphi\Bigl(\!\displaystyle\frac{\Delta\vec
x}{\sqrt{D}},t,D\!\Bigr)
\exp\Bigl[\!\displaystyle\frac{1}{D}S(t,D)\Bigr]\!\!\biggr\}.
 \label{fkpp4}
 \end{equation}
  Here $\Phi(\vec x, t, D)$ is a common element of the class, the real function $\varphi(\vec{\xi}, t, D)$ belongs to the Schwartz space $\mathbb{S}$ in variables $\vec{\xi}$, smoothly depends on $t$ and regularly depends on $\sqrt{D}$ as  $D\to 0$, $\Delta\vec{x} = \vec{x} - \vec{X} (t, D)$. The real smooth functions $\vec{X}(t, D)$ and $S(t, D)$, characterizing the class ${\mathcal P}_D^t$, regularly depend on $\sqrt{D}$ as $D\to 0$ and are to be determined when constructing a solution to the equation \eqref{fkpp3}.

 The functions of the class ${\mathcal P}_D^t$ are concentrated, as $D\to0$, in a neighborhood
of a point moving in the coordinate space along a curve given by the equation $\vec{x}=\vec{X}(t,0)$.

 In addition to $\Delta\vec{x}$, we introduce the following operators acting on functions of the class ${\mathcal P}_D^t$:
 $\hat{\vec p}=D\partial_{\vec x}$, $\hat T=D\partial_t+\langle\dot{\vec X}(t,D),\hat{\vec p}\rangle-\dot{S}(t,D)$,
  and $\hat{\Delta}_{\alpha,\beta}(t,D)$,
    where $\langle,\rangle$ means the scalar product of $n$--dimensional vectors; 
    $\partial_{\vec x}$ is the gradient operator in Cartesian coordinates $\vec{x}$; the operator $\hat{\Delta}_{\mu,\nu}(t,D)$ is defined by its Weil symbol ${\Delta}_{\mu,\nu}(\vec{p},\vec{x},t,D)=\vec{p}^\mu \Delta\vec{x}^\nu$,  $\vec{p}\in{\mathbb{R}}^n$ is the symbol of the operator $\hat{\vec p}$;     $\mu, \nu \in{\mathbb{Z}}_{+}^n$ are multi-indices:
    \begin{align}
  &\nu=(\nu_1,\nu_2, \ldots , \nu_n),\quad \nu_1, \nu_1, \ldots , \nu_n\in {\mathbb{Z}}_{+}^1,  \cr
   & |\nu|=\nu_1+ \nu_2+\ldots +\nu_n,\quad \nu !=\nu_1!\nu_2!\ldots \nu_n!.
    \label{multind}
    \end{align}
For any vector $\vec{a}=(a_1,a_2, \ldots ,a_n)\in {\mathbb{R}}^n$ we denote $\vec{a}^\nu= (a_1^{\nu_1},a_2^{\nu_2}, \dots , a_n^{\nu_n} )\in{\mathbb{R}}^n $.

One can directly verify the validity of the following asymptotic estimates for the operators and the functions from the class
${\mathcal P}_D^t$ \cite{bagrov1,LST16}:
\begin{align}
&\displaystyle\frac{\parallel \hat{\Delta}_{\mu,\nu}(t,D)\Phi \parallel}{\parallel \Phi \parallel}={\rm O}(D^{(|\mu|+|\nu |)/2}), \quad
\displaystyle\frac{\parallel \hat{T}\Phi \parallel}{\parallel \Phi \parallel}={\rm O}(D),\quad \Phi\in {\mathcal P}_D^t,
\label{estim}
\end{align}
where the norm $\parallel \ldots  \parallel$  is meant in the sense of the space $L_2$.

Formulas \eqref{estim} can be considered as estimates of the operators acting on functions of the class ${\mathcal P}_D^t$:
\begin{align}
& \hat{\Delta}_{\mu,\nu}(t,D)= {\rm \hat{O}}(D^{(|\mu|+|\nu |)/2}), \quad
\hat{T}={\rm \hat{O}}(D),
\label{estim1}
\end{align}
and, in particular, $\Delta \vec{x} = {\rm \hat{O}}(\sqrt{D})$, $\hat{\vec{p}} ={\rm \hat{O}}(\sqrt{D})$.
Here, ${\rm \hat{O}}({D}^{k})$, $k\geq 0$,  means an operator $\hat{F}$ such that
${\parallel \hat{F}\phi \parallel}/{\parallel \phi \parallel}={\rm O}(D^k)$, $\phi\in{\mathcal P}_D^t$.

In the asymptotic estimates,  the leading term gives more insight into the solution of the  problem. Therefore, in this work we focus on constructing the leading terms of asymptotic solutions of the kinetic equation \eqref{fkpp3}. In the WKB-Maslov theory, the key object of the asymptotic approach is the semiclassical closed dynamical system describing moments of the unknown solution to the nonlinear equation. Such the system will be obtained in the next section.

\section{The Einstein--Ehrenfest system of the second order}
\label{sec:3ee}

The  semiclassical approach developed in \cite{fkppshap18,shapovalov:BTS1,sym2020} can be applied to equation \eqref{fkpp3} when
the following moments exist for its solution  $u(\vec{x},t,D)\in{\mathcal{P}}_D^t$:
\begin{align}
&\sigma_u (t,D)=\int\limits_{{\mathbb{R}}^n}u(\vec{x},t,D)d\vec{x}, \quad
\vec{x}_u(t,D)=\displaystyle\frac{1}{\sigma_u(t,D)}\int\limits_{{\mathbb{R}}^n}\vec{x}u(\vec{x},t,D)d\vec{x} \label{moment1a}\\
& \alpha^{\nu}_{u}(t,D)=\displaystyle\frac{1}{\sigma_u(t,D)}\int\limits_{{\mathbb{R}}^n} \Delta\vec{x}\,^\nu
u(\vec{x},t,D) d\vec{x},
\quad \nu=(\nu_1,\nu_2,\ldots ,\nu_n)\in {\mathbb{Z}}_{+}^n,\,\, |\nu|\in {\mathbb{Z}}_{+}^1.
\label{moment1b}
\end{align}
Here, the zeroth-order  moment $\sigma_u (t,D)$ has the meaning of the number of ions in the plasma at  time $t$.

We choose  the vector $\vec{X}(t,D)$ characterising the class ${\mathcal{P}}_D^t$ to be equal to the first normalized moment
\begin{align}
\vec{X}(t,D)=\vec{x}_u(t,D).
\label{moment2}
\end{align}
Then $\alpha^{\nu}_{u}(t,D)=0$ for $|\nu|=1$.

We also limit our consideration to solutions $u(\vec{x},t,D)$ of \eqref{fkpp3} with  $\sigma_u(t,D)= {\rm O}(1)$. Otherwise, the nonlinear term would be infinitely large compared to the linear term as $D\to 0$, i.e. the rate of triple recombination would dominate over the ionization rate at each $t$. In \eqref{fkpp2} the $\varkappa$ is defined so that the ionization processes compete with the triple recombination that is the most interesting case from a physical point of view.

Then from \eqref{estim} we have
\begin{align}
&\vec{x}_u(t,D)= {\rm O}(1), \quad \alpha^{\nu}_{u}(t,D)= {\rm O}(D^{|\nu|/2}).
\label{moment3}
\end{align}

For constructing  the leading term of the semiclassical asymptotic solution to equation \eqref{fkpp3}, we consider a set of moments of the form \eqref{moment1b}, including  $\sigma_u(t,D)$, $\vec{x}_u(t,D)$, and the second-order moments $\alpha^{\nu}_{u}(t,D)$, $|\nu|=2$,  which  can be represented in the form of a $n$--dimensional symmetric matrix
\begin{align}
&\alpha^{(2)}_u(t,D)= \big(\alpha^{(2)}_{u,ij}(t,D) \big),\quad
\label{moment2a}
\end{align}
where $\alpha^{(2)}_{u,ij}(t,D)=\displaystyle\frac{1}{\sigma_u(t,D)}\displaystyle\int\limits_{{\mathbb{R}}^n}\Delta x_i \Delta x_j u(\vec{x},t,D) d\vec{x}$, and $i,j=1,2,\ldots ,n$.

For simplicity of notation, we introduce an aggregate vector of the considered moments
\begin{align}
&\Theta_u(t,D)=\big(\sigma_u(t,D), \vec{x}_u(t,D), \alpha^{(2)}_u(t,D)  \big).
\label{moment4}
\end{align}

In what follows, we will omit the function arguments, including the asymptotic parameter $D$, if this does not lead to a misunderstanding.

Let us obtain the dynamical system describing evolution of the moments \eqref{moment4}.
To do this, we represent the functions $a(\vec{x},t)$  and $b(\vec{x},\vec{y},\vec{z},t)$ in equation \eqref{fkpp3} as the  second-order Taylor series expansions  about the point $\vec{X}(t,D)$.
Using  matrix notations, we can write
\begin{align}
&a(\vec{x},t)=a(\vec{X},t)+a_{x}\Delta\vec{x}+\frac{1}{2}\Delta\vec{x}^Ta_{xx}\Delta\vec{x}+\ldots\,\,(\mbox{higher order terms}),
\label{taylor1}
\end{align}
\begin{align}
&b(\vec{x},\vec{y},\vec{z},t)=b(\vec{X},\vec{X},\vec{X},t)+b_{x}\Delta\vec{x}+b_{y}\Delta\vec{y}+b_{z}\Delta\vec{z}+
\frac{1}{2}\Big(\Delta\vec{x}^T b_{xx}\Delta\vec{x}+\cr
&+\Delta\vec{x}^T b_{xy}\Delta\vec{y}+ \Delta\vec{y}^T b_{yx}\Delta\vec{x}+
\Delta\vec{x}^T b_{xz}\Delta\vec{z}+ \Delta\vec{z}^T b_{zx}\Delta\vec{x}+ \Delta\vec{y}^T b_{yy}\Delta\vec{y}+\cr
&\Delta\vec{y}^T b_{yz}\Delta\vec{z}+ \Delta\vec{z}^T b_{zy}\Delta\vec{y}+
\Delta\vec{z}^T b_{zz}\Delta\vec{z}\Big)+\ldots\,\,(\mbox{higher order terms}).
\label{taylor2}
\end{align}
Here, $\Delta\vec{x}=\vec{x}-\vec{X}(t)$, $\Delta\vec{y}=\vec{y}-\vec{X}(t)$, $\Delta\vec{z}=\vec{z}-\vec{X}(t)$ are column vectors, the transposition is indicated by $T$; $a_x$, $b_x$, $b_y$, $b_z$ are row vectors, $a_x=\big(\displaystyle\frac{\partial a}{\partial x_i}|_{\vec{x}=\vec{X}(t)}\big)$, $b_x=\big(\displaystyle\frac{\partial b}{\partial x_i}|_{\vec{x}=\vec{X}(t),\vec{y}=\vec{X}(t),\vec{z}=\vec{X}(t)}\big)$, and the same for $b_y$ and $b_z$;
$a_{xx}$, $b_{xx}$, $b_{yy}$, $b_{zz}$, $b_{xy}=b_{yx}$, $b_{xz}=b_{zx}$, $b_{zy}=b_{yz}$ are $n$-dimensional matrices of the form
$a_{xx}=\big(\displaystyle\frac{\partial^2 a}{\partial x_i\partial x_j}|_{\vec{x}=\vec{X}(t)}\big)$,
$b_{xx}=\big(\displaystyle\frac{\partial^2 b}{\partial x_i\partial x_j}|_{\vec{x}=\vec{X}(t),\vec{y}=\vec{X}(t),\vec{z}=\vec{X}(t)}\big)$,
$b_{xy}=\big(\displaystyle\frac{\partial^2 b}{\partial x_i\partial y_j}|_{\vec{x}=\vec{X}(t),\vec{y}=\vec{X}(t),\vec{z}=\vec{X}(t)}\big)$, and the same for $b_{yy}$, $b_{zz}$, $b_{xz}=b_{zx}$, $b_{zy}=b_{yz}$.

We also consider a particular expansion
\begin{align}
&b(\vec{x},\vec{y},\vec{z},t)=b(\vec{x},\vec{X},\vec{X},t)+b_{y}(\vec{x})\Delta\vec{y}+b_{z}(\vec{x})\Delta\vec{z}+
\frac{1}{2}\Big(\Delta\vec{y}^T b_{yy}(\vec{x})\Delta\vec{y}+\cr
&\Delta\vec{y}^T b_{yz}(\vec{x})\Delta\vec{z}+ \Delta\vec{z}^T b_{zy}(\vec{x})\Delta\vec{y}+
\Delta\vec{z}^T b_{zz}(\vec{x})\Delta\vec{z}\Big)+\ldots\,\,(\mbox{higher order terms}),
\label{taylor2a}
\end{align}
where $b_{y}(\vec{x})=$ $b_y(\vec{x},\vec{X},\vec{X})$= $\big(\displaystyle\frac{\partial b}{\partial y_i}|_{\vec{y}=\vec{X}(t),\vec{z}=\vec{X}(t)}\big)$ and the same for $b_{z}(\vec{x})$;
$b_{yy}(\vec{x})=$ $b_{yy}(\vec{x},\vec{X}, \vec{X})$= $\big(\displaystyle\frac{\partial^2 b}{\partial y_i\partial y_j}|_{\vec{y}=\vec{X}(t),\vec{z}=\vec{X}(t)}\big)$,
and the same for $b_{yz}(\vec{x})=b_{zy}(\vec{x})$, and $b_{zz}(\vec{x})$.

To derive a dynamical system for moments \eqref{moment4},
we differentiate the moments \eqref{moment4} with respect to time  and substitute the derivative $\partial_t u$ from equation \eqref{fkpp3}. Taking into account expansions \eqref{taylor1} and \eqref{taylor2}, and keeping  the expansion terms no higher than the second order, we arrive at the equations
\begin{align}
&\dot{\sigma}_u=\sigma_u\Big(a(\vec{X},t)+\displaystyle\frac{1}{2}{\rm Sp}\big[a_{xx} \alpha^{(2)}_u \big] \Big)- \varkappa\sigma_u^3\Big( b(\vec{X},\vec{X},\vec{X},t)+ \cr
&+\displaystyle\frac{1}{2}{\rm Sp}\big[(b_{xx}+b_{yy}+b_{zz})\alpha^{(2)}_u \big] \Big), \label{moment5a}\\
&\dot{\vec{x}}_u=\big(a_x-\varkappa \sigma_u^2b_x\big) \alpha^{(2)}_u, \label{moment5b}\\
&\dot{\alpha}^{(2)}_u=2D \tilde{D}_a(t){\mathbb{I}}_{n\times n}.
\label{moment5c}
\end{align}
Here, dot denotes the time derivative (e.g., $\dot{\sigma}_u=d\sigma_u/dt$),  $a_{xx}\alpha^{(2)}_u$  means matrix product,
 and ${\mathbb{I}}_{n\times n}$ is the identity matrix of size $n$, $\vec{X}=\vec{x}_u(t,D)$ according to \eqref{moment2}.

We can also rewrite equations \eqref{moment5a} --- \eqref{moment5c} more succinctly using the aggregate vector \eqref{moment4} as
\begin{align}
&\dot{\Theta}_u(t,D)=\Gamma(\Theta_u(t,D),t,D),
\label{moment6a}
\end{align}
where
\begin{align}
&\Gamma(\Theta_u(t,D),t,D)=\big(f(\Theta_u(t,D),t,D), g(\Theta_u(t,D),t,D), h(\Theta_u(t,D),t,D)\big),
\label{moment6b}
\end{align}
consists of the functions $f$, $g$, $h$ corresponding to  $\dot{\sigma}_u$, $\dot{\vec{x}}_u$, and  $\dot{\alpha}^{(2)}_u$, the
form of which is obvious from \eqref{moment5a}, \eqref{moment5b}, and \eqref{moment5c}, respectively.

Consider now a system of ordinary differential equations (ODEs)
\begin{align}
&\dot{\Theta}(t)=\Gamma(\Theta(t),t,D)
\label{moment7}
\end{align}
 for an aggregate vector
 \begin{align}
&\Theta(t)=\big(\sigma(t), \vec{x}(t), \alpha^{(2)}(t)  \big).
\label{moment8}
\end{align}
Here, $\Gamma(\Theta(t),t,D)$ is taken from \eqref{moment6b};  real variable $\sigma\in {\mathbb{R}}^1$, real vector $\vec{x}\in {\mathbb{R}}^n$,  and real symmetric matrix  $\alpha^{(2)}=(\alpha^{(2)}_{ij})$ are the dependent variables.

Further in consideration of the system \eqref{moment7}, we follow the papers \cite{shap2009,LST14,fkppshap18}.

According to \cite{shap2009,LST14,fkppshap18}, we call equations \eqref{moment7} the {\it Einstein--Ehrenfest (EE) system} of the second order for the kinetic equation \eqref{fkpp3}. The second order of the EE system means the presence of $\alpha^{(2)}$ in \eqref{moment8}.

Note that the Cauchy problem for the EE system \eqref{moment7} is known to have a unique solution under some conditions on the coefficients of the system, which are assumed to be satisfied.

Let equation \eqref{fkpp3} have a solution $u(\vec{x},t,D)$ belonging to the class ${\mathcal P}_D^t$ given by \eqref{fkpp4} with the initial function
\begin{align}
&u(\vec{x},t,D)|_{t=0}=\varphi(\vec{x},D),
\label{cauchy1}
\end{align}
where the function $\varphi(\vec{x}, D) $ belongs to the class ${\mathcal P}_D^0$ of trajectory-concentrated functions \eqref{fkpp4} for $t=0$, $ {\mathcal P}_D^0={\mathcal P}_D^t|_{t=0}$.
Then we can set the initial conditions for the the EE system \eqref{moment7}  as
\begin{align}
&\Theta(t)|_{t=0}= {\mathfrak{g}}_\varphi=\big(\sigma_\varphi, \vec{x}_{\varphi}, \alpha_\varphi ^{(2)} \big),
\label{cauchy-ee}
\end{align}
where
\begin{align}
&\sigma|_{t=0}=\sigma_\varphi =\displaystyle\int\limits_{{\mathbb{R}}^n}\varphi(\vec{x},D)d\vec{x},\quad
 \vec{x}|_{t=0}=\vec{x}_\varphi=\frac{1}{\sigma_\varphi}\displaystyle\int\limits_{{\mathbb{R}}^n}\vec{x}\varphi(\vec{x},D)d\vec{x},\cr
&\alpha^{(2)}_{ij} |_{t=0}=
\alpha_{\varphi,ij}^{(2)}=\frac{1}{\sigma_\varphi}\displaystyle\int\limits_{{\mathbb{R}}^n}
(x_i-x_{\varphi,i})(x_j-x_{\varphi,j})\varphi(\vec{x},D)d\vec{x}.
\label{cauchy-ee1}
\end{align}
We write the solution of the Cauchy problem for the EE system \eqref{moment7}  with the initial conditions \eqref{cauchy-ee},  \eqref{cauchy-ee1} in the form
\begin{align}
{\mathfrak{g}}_\varphi(t)=\big(\sigma_\varphi(t), \vec{x}_{\varphi}(t), \alpha_\varphi ^{(2)}(t) \big),
\label{cauchy-ee2}
\end{align}
and
\begin{equation}
\label{gensol}
{\mathfrak{g}}(t,{\bf C})=\big(\sigma(t,{\bf C}), \vec{x}(t,{\bf C}),\alpha^{(2)}(t,{\bf C})\big)
\end{equation}
is the general solution of the EE system \eqref{moment7}, where $\bf C$ is a set of arbitrary integration constants.
We omit the argument $D$ in $\sigma_{\varphi}(t) $, $\vec{x}_\varphi (t)$, and $\alpha_\varphi^{(2)}(t)$ for brevity.

Denote by   ${\bf C}_\varphi$ the solution of the following {\it algebraic} equation involving arbitrary integration constants  ${\bf C}$ as unknowns:
\begin{equation}
\label{alg}
{\mathfrak{g}}(0,{\bf C})={\mathfrak{g}}_\varphi,
\end{equation}
i.e., ${\mathfrak{g}}(0,{\bf C}_\varphi)={\mathfrak{g}}_\varphi$.

We consider the solution of equation \eqref{alg} in specific examples, leaving aside the general algebraic problem of solvability of this equation.
Note also that ${\bf C}_\varphi$ is a vector functional of $\varphi$.

From the  uniqueness of the solution of the Cauchy problem \eqref{moment7}, \eqref{moment8}, in view of \eqref{cauchy-ee}, \eqref{cauchy-ee1} and the condition \eqref{alg}, we have \cite{shap2009,fkppshap18}
\begin{equation}
{\mathfrak{g}}(t,{\bf C}_\varphi)={\mathfrak{g}}_\varphi (t).
\label{cauchy-ee3}
\end{equation}
On the other hand, we can consider the aggregate vector of moments $\Theta_u(t,D)$ \eqref{moment4} being determined by the solution
$u(\vec{x},t,D)$  of equation \eqref{fkpp3} with the initial condition \eqref{cauchy1}. For uniformity, we denote it by
\begin{align}
{\mathfrak{g}}_u(t)=\big(\sigma_u(t), \vec{x}_u(t),\alpha_u^{(2)}(t)\big).
\label{cauchy-ee4}
\end{align}

Since we  consider the EE system of the second order, then $u(\vec{x},t,D)$ should be a leading term
of the semiclassical asymptotic to equation \eqref{fkpp3} accurate to ${\rm O}(D^{3/2})$ and the subsequent analysis is carried out with this accuracy.

It can be seen that ${\mathfrak{g}}_u(t)$ satisfies the EE system \eqref{moment7} and the initial condition \eqref{cauchy-ee}
\begin{align}
{\mathfrak{g}}_u(0)={\mathfrak{g}}_\varphi.
\label{cauchy-ee5}
\end{align}
Then we have
\begin{align}
{\mathfrak{g}}_u(t)={\mathfrak{g}}_\varphi (t).
\label{cauchy-ee6}
\end{align}

 In conclusion of this section, we defines values ${\bf C}_u(t)$ from the algebraic condition \eqref{alg} but taken for any instant $t$:
 \begin{equation}
\label{alg1}
{\mathfrak{g}}(t,{\bf C}_u(t))={\mathfrak{g}}_\varphi(t).
\end{equation}
 From \eqref{alg1}, \eqref{cauchy-ee6}, and \eqref{cauchy-ee3}, we find that ${\bf C}_u(t)={\bf C}_\varphi$, i.e.
 the  functionals ${\bf C}_u(t)$  can be considered as approximate integrals for the equation \eqref{fkpp3} in the class
 \eqref{fkpp4} accurate to ${\rm O}(D^{3/2})$. In the next section, we will show that solutions of the dynamical system \eqref{moment5c} generate a family of auxiliary linear equations associated with the nonlinear kinetic equation, and the algebraic condition \eqref{alg} allows one to find the asymptotic solutions of the nonlinear kinetic equation among solutions of these linear equations.

\section{Auxiliary linear problem and the Cauchy problem}
\label{sec:linprob}

For constructing the leading term of the semiclassical asymptotic solution in the class \eqref{fkpp4}, we first substitute the expansion \eqref{taylor2a} into the equation \eqref{fkpp3}. In view of the estimates \eqref{estim} and formulae for the moments
\eqref{moment1a}, \eqref{moment1b}, \eqref{moment2a}, we write
\begin{equation}
\begin{gathered}
\Bigg\{-\partial_t + D \tilde{D}_a(t)\Delta  + a(\vec{x},t)-\varkappa \sigma^2_u(t) \bigg(b(\vec{x},\vec{X}(t),\vec{X}(t),t)+ \cr
+\displaystyle\frac{1}{2} {\rm Sp}\Big[\big(b_{yy}(\vec{x},\vec{X},\vec{X},t)+b_{zz}(\vec{x},\vec{X},\vec{X},t)\big) \alpha^{(2)}_u(t)\Big] \bigg) \Bigg\}u(\vec{x},t)={\rm O}(D^{3/2}).
\end{gathered}
\label{fkpp9}
\end{equation}
Here, the evolution of the moments $\sigma_u(t)$, $\vec{X}=\vec{x}_u(t,D)$, and $\alpha^{(2)}_u(t)$ is governed by the
dynamical system of the second order  \eqref{moment5a} --- \eqref{moment5c} or \eqref{moment6a}, \eqref{moment6b}
with the initial condition \eqref{cauchy-ee} when the initial condition \eqref{cauchy1} holds.

Next, we replace the moments \eqref{cauchy-ee4}  in the equation \eqref{fkpp9} with the general solution of EE systems of the second order of the form \eqref{moment7}, \eqref{moment8} given by \eqref{gensol} and go over the next linear equation parametrized by the arbitrary integration constants ${\bf C}$:
\begin{align}
\hat{L}(\vec{x},t,{\bf C})v(\vec{x},t)=0,
\label{lineq1}
\end{align}
where
\begin{align}
&\hat{L}(\vec{x},t,{\bf C})=-\partial_t+D \tilde{D}_a(t)\Delta +a(\vec{x},t)-\varkappa\sigma^2(t,{\bf C})
\bigg( b\big(\vec{x}, \vec{x}(t,{\bf C}),\vec{x}(t,{\bf C})\big)+ \cr
&+\frac{1}{2}{\rm Sp}\big[\big( b_{yy}(\vec{x},\vec{x}(t,{\bf C}),\vec{x}(t,{\bf C}))+
b_{zz}(\vec{x},\vec{x}(t,{\bf C}),\vec{x}(t,{\bf C})) \big) \alpha^{(2)}(t,{\bf C}) \big] \bigg).
\label{lineq1a}
\end{align}
Here,  $b_{yy}(\vec{x},\vec{x}(t,{\bf C}), \vec{x}(t,{\bf C}))=$
 $\big(\displaystyle\frac{\partial^2 b}{\partial y_i\partial y_j}|_{\vec{y}=\vec{x}(t,{\bf C}),\vec{z}=\vec{x}(t,{\bf C})}\big)$,
and the same for $b_{zz}$.

By analogy, we can construct the following linear equation from  \eqref{fkpp3} with the use of expansions
\eqref{taylor1} and \eqref{taylor2}:
\begin{align}
\hat{{\mathcal{L}}}(\vec{x},t,{\bf C})v(\vec{x},t)=0,
\label{lineq2}
\end{align}
where
\begin{align}
&\hat{{\mathcal{L}}}(\vec{x},t,{\bf C})=-\partial_t+D \tilde{D}_a(t)\Delta + L(t,{\bf C})+
L_{x}(t,{\bf C})\Delta\vec{x}+\displaystyle\frac{1}{2}\Delta\vec{x}^T L_{xx}(t,{\bf C})\Delta\vec{x},
\label{lineq2a}
\end{align}
and
\begin{align}
&L(t,{\bf C})= a(\vec{x}(t,{\bf C}),t)-\varkappa\sigma^2(t,{\bf C})
\bigg( b\big(\vec{x}(t,{\bf C}), \vec{x}(t,{\bf C}),\vec{x}(t,{\bf C})\big)+ \cr
&+\frac{1}{2}{\rm Sp}\big[\big( b_{yy}(\vec{x}(t,{\bf C}),\vec{x}(t,{\bf C}),\vec{x}(t,{\bf C}))+
b_{zz}(\vec{x}(t,{\bf C}),\vec{x}(t,{\bf C}),\vec{x}(t,{\bf C})) \big) \alpha^{(2)}(t,{\bf C}) \big] \bigg),\cr
&L_{x}(t,{\bf C})= a_{x}(\vec{x}(t,{\bf C}),t)-\varkappa\sigma^2(t,{\bf C})b_{x}\big(\vec{x}(t,{\bf C}), \vec{x}(t,{\bf C}),\vec{x}(t,{\bf C})\big), \cr
& L_{xx}(t,{\bf C})=a_{xx}(\vec{x}(t,{\bf C}),t)-\varkappa\sigma^2(t,{\bf C}) b_{xx}\big(\vec{x}(t,{\bf C}), \vec{x}(t,{\bf C}),\vec{x}(t,{\bf C})\big).
\label{lineq2b}
\end{align}
Note that in view of estimates \eqref{estim} and \eqref{estim1}, we can see that
\begin{align}
&\hat{L}(\vec{x},t,{\bf C})=\hat{{\mathcal{L}}}(\vec{x},t,{\bf C})+{\rm \hat{O}}(D^{3/2}).
\label{lineq3}
\end{align}

By analogy with \cite{shap2009,LST14,fkppshap18}, we use the term {\it associated linear equation} (ALE) for \eqref{lineq2} with the coefficients \eqref{lineq2a} and \eqref{lineq2b}.

Following the Maslov method \cite{Maslov1}, we need the operator \eqref{lineq1a} to satisfy
\begin{equation}
\hat{L}(\vec{x},t,{\bf C})={\rm \hat{O}}(1),
\label{lcond1}
\end{equation}
so that the equation \eqref{lineq1} determines the leading term of asymptotics in the class \eqref{fkpp4}. In view of estimates \eqref{estim1}, the condition \eqref{lcond1} is satisfied if the free function $S(t,D)$ characterizing the class \eqref{fkpp4} has the estimate $\dot{S}(t,D)={\rm O}(D)$. Without loss of generality, we choose it in the following form
\begin{equation}
S(t,D,{\bf C})=D L(t,{\bf C}).
\label{sdef1}
\end{equation}
It can be shown that function $S(t,D,{\bf C})$ defined by \eqref{sdef1} satisfies
\begin{equation}
\exp\Big[\dac{1}{D}S(t,D,{\bf C})\Big]=\dac{\sigma(t,{\bf C})}{\sigma(0,{\bf C})}+{\rm O}(D).
\label{sdef2}
\end{equation}

Consider the Cauchy problem for equation \eqref{lineq1} or \eqref{lineq2} with the initial condition \eqref{cauchy1} supposing
\begin{align}
v(\vec{x},t)|_{t=0}=\varphi(\vec{x},D), \quad \varphi(\vec{x},D)\in {\mathcal P}_D^0.
\label{cauchylin1}
\end{align}
Replace the arbitrary constants ${\bf C}$ in equation  \eqref{lineq1} or \eqref{lineq2} by the constants ${\bf C}_\varphi$
determined by the algebraic condition \eqref{alg}. Considering \eqref{cauchy-ee3}, \eqref{cauchy-ee6}, and \eqref{lineq3}, we can see that equation \eqref{lineq1} or \eqref{lineq2} transforms into equation \eqref{fkpp9} accurate to ${\rm O}(D^{3/2})$.
Then the following theorem holds \cite{shap2009,LST14,fkppshap18}.

\begin{theorem} The solutions of the Cauchy problem for the nonlinear equation \eqref{fkpp9} and
of the Cauchy problem for the associated linear equation \eqref{lineq1} or \eqref{lineq2} with the same initial condition
\eqref{cauchy1} and \eqref{cauchylin1} are related as
\begin{align}
u(\vec{x}, t) = v(\vec{x}, t,{\bf C}_\varphi) + {\rm O}(D^{3/2}),
\label{cauchylin2}
\end{align}
where the constants ${\bf C}_\varphi$ are determined by the algebraic condition \eqref{alg}.
\end{theorem}

The forms of the associated linear equation operator related by \eqref{lineq3} are termed equivalent \cite{Maslov1}. The form \eqref{lineq2a} can be more profitable to construct solutions $v(\vec{x},t)$. In particular, the Green function can be obtained in the explicit form for the equation \eqref{lineq2}, which is quadratic in $\vec{x}$, using the Fourier transform in a similar way as it was done in \cite{bagrov1}. The expression for the Green function of \eqref{lineq2} is cumbersome in a general case so we confine ourselves to the construction of the evolution operator just for the special case considered in the next section.

Let us note one more important fact. In notations \eqref{lineq2b}, the Cauchy problem for the system \eqref{moment5a}, \eqref{moment5b}, \eqref{moment5c} reads
\begin{equation}
\begin{gathered}
\dot{\sigma}=\sigma\bigg( L(t,{\bf C}) + \displaystyle\frac{1}{2}\Sp\Big[L_{xx}(t,{\bf C})\cdot \alpha^{(2)}\Big]\bigg), \\
\dot{\vec{X}}=L_{\vec{x}}(t,{\bf C})\cdot \alpha^{(2)}, \\
\dot{\alpha}^{(2)}=2D\tilde{D}_a(t){\mathbb{I}}_{n\times n}, \\
{\bf C}=\Big(\sigma(t),\vec{X}(t),\alpha^{(2)}(t)\Big)\Big|_{t=0}.
\end{gathered}
\label{fkpp12}
\end{equation}
Therefore, the moments of the function $u(\vec{x},t)$ are determined by the leading term of its asymptotics, $v(\vec{x},t,{\bf C}_{\varphi})$, within the accuracy of ${\rm O}(D^{3/2})$.

Next section illustrates the formalism of our approach with the specific example.

\begin{section}{Plasma relaxation}
\label{sec:example}
In this section, we consider the example of application of our method to the equation \eqref{fkpp2}, \eqref{fkpp3} that describes the relaxation of the plasma with the uniformly heated atom and electron gases, i.e. the case $T_g=\rm{const}$, $T_e(\vec{x},t)=T_e(t)$. Since the ion concentration is localized on the axis of the GDT in the metal vapor active media, the two-dimensional problem in the GDT cross-section is considered ($\vec{x}=(x_1,x_2)$). It is assumed that the neutral atoms concentration is almost independent of spatial coordinates and of the time ($n_{neut}(\vec{x},t)=\rm{const}$). The independence of $T_e$ from $\vec{x}$ yields $a(\vec{x},t)=\tilde{a}(t)$, $b(\vec{x},\vec{y},\vec{z},t)=\tilde{b}(\vec{x}-\vec{y},\vec{x}-\vec{z},t)$ in view of \eqref{fkpp2a}. The relaxation process implies the monotone decrease in the electron temperature over time. In such the process, the function $\tilde{a}(t)$ monotonically decrease and $\tilde{b}(\vec{x}-\vec{y},\vec{x}-\vec{z},t)$ monotonically increase over time.

In view of our assumptions, we have $\dot{\vec{X}}=0$ from \eqref{moment5b}. Let the GDT axis be the origin of coordinates and the initial distribution of the ions be axially symmetric. Then, \eqref{moment5b} and \eqref{moment5c} read
\begin{equation}
\vec{X}(t)=0, \qquad \alpha^{(2)}(t)={\mathbb{I}}_{2\times 2}\Big[D_{in}+2D\displaystyle\int\limits_{0}^{t}\tilde{D}_a(\tau)d\tau\Big],
\label{fkprim1}
\end{equation}
where $\alpha^{(2)}(0)=D_{in}{\mathbb{I}}_{2\times 2}$ and $D_{in}={\rm O}(D)$ is the coefficient that determines the initial localization area of ions. The nonlocality kernel is approximated by the delta-like function of the following form:
\begin{equation}
\tilde{b}(\vec{r}_1,\vec{r}_2,t)=\beta(t)\exp\Big[-\displaystyle\frac{\vec{r}_1\,^2+\vec{r}_2\,^2}{2\varrho^2}\Big].
\label{fkprim2}
\end{equation}
The substitution of \eqref{fkprim1} and \eqref{fkprim2} into \eqref{moment5a} yields
\begin{equation}
\dot{\sigma}=\sigma \tilde{a}(t)-\varkappa \sigma^3 \beta(t) \bigg[1-4\displaystyle\frac{D_{in}}{\varrho^2}-8\displaystyle\frac{D}{\varrho^2}\displaystyle\int\limits_{0}^{t}\tilde{D}_a(\tau)d\tau\bigg]
\label{fkprim3}
\end{equation}
The equation \eqref{fkprim3} is the Bernoulli differential equation. We search its solutions in the form
\begin{equation}
\sigma(t)=U(t)\cdot \exp \Big[\displaystyle\int\limits_{0}^t \tilde{a}(\tau)d\tau\Big].
\label{fkprim4}
\end{equation}
Then, we have
\begin{equation}
U(t)=\Bigg[\displaystyle\int\limits_{0}^{t}\Bigg(2\varkappa \exp \Big[2\displaystyle\int\limits_{0}^\theta \tilde{a}(\tau)d\tau\Big] \beta(\theta) \bigg[1-4\displaystyle\frac{D_{in}}{\varrho^2}-8\displaystyle\frac{D}{\varrho^2}\displaystyle\int\limits_{0}^{\theta}\tilde{D}_a(\tau)d\tau\bigg]\Bigg)d\theta+\displaystyle\frac{1}{\sigma^2(0)}\Bigg]^{-1/2}.
\label{fkprim5}
\end{equation}
Let us state the minimum restrictions for functions $\tilde{a}(t)$, $\beta(t)$, $\tilde{D}_a(t)$ so that the meet the physical meaning of the problem at an arbitrary time interval. Since $\tilde{D}_a(t)\sim\Big(1+\displaystyle\frac{T_e(t)}{T_g}\Big)$, the function $\tilde{D}_a(t)$ must be a decreasing function. Also, the conditions $\tilde{a}(t)>0$ and $\beta(t)>0$ must be met due to the non-negativity of the probability of ionization and triple recombination acts. Moreover, the function $\beta(t)$ must be bounded above as well as the electron temperature. Also, we have already mentioned that $\tilde{a}(t)$ must be a decreasing function and $\beta(t)$ must be an increasing one. Finally, we assume the processes to be exponential, that is the simple approximation often used in various problems, and functions $\tilde{a}(t)$, $\beta(t)$, $\tilde{D}_a(t)$ read
\begin{equation}
\tilde{a}(t)=A_1 e^{-t/\tau_a}, \qquad \tilde{D}_a(t)=d_1 e^{-t/\tau_d}, \qquad \beta(t)=B_2+(B_1-B_2) e^{-t/\tau_b}.
\label{fkprim5a}
\end{equation}
Here, $\tau_a$, $\tau_d$, $\tau_b$ are time constants for the change over time of the ionization rate, the ambipolar diffusion and the triple recombination rate respectively, $A_1$ is the initial ionization rate, $d_1$ is proportional the initial ambipolar diffusion coefficient, coefficients $B_1$ and $B_2$ are proportional to initial and final triple recombination rates respectively. Then, we have
\begin{equation}
\sigma(t)=\exp\Big[-A_1 \tau_a e^{-t/\tau_a}+A_1\tau_a\Big]\cdot \Bigg[\displaystyle\frac{2\varkappa}{\varrho^2}F(t)+\displaystyle\frac{1}{\sigma^2(0)}\Bigg]^{-1/2},
\label{fkprim6}
\end{equation}
where
\begin{equation}
\begin{gathered}
F(t)=\tau_a e^{2 A_1 \tau_a} \bigg\{ (\varrho^2-4D_{in}-8Dd_1\tau_d )B_2  \Gamma\Big[0,2A_1\tau_a e^{-t/\tau_a},2A_1\tau_a\Big]+ \\
+(\varrho^2-4D_{in}-8D d_1 \tau_d)(B_1-B_2) (2A_1\tau_a)^{-\tau_a/\tau_b}\Gamma\Big[\frac{\tau_a}{\tau_b},2A_1\tau_a e^{-t/\tau_a},2A_1\tau_a\Big]- \\
-8D  d_1\tau_d B_2  (2A_1\tau_a)^{-\tau_a/\tau_d}\Gamma\Big[\frac{\tau_a}{\tau_d},2A_1\tau_a e^{-t/\tau_a},2A_1\tau_a\Big]- \\
-8D  d_1 \tau_d (B_1-B_2) (2A_1\tau_a)^{-\frac{\tau_a(\tau_b+\tau_d)}{\tau_b \tau_d}}\Gamma\Big[\frac{\tau_a(\tau_b+\tau_d)}{\tau_b \tau_d},2A_1\tau_a e^{-t/\tau_a},2A_1\tau_a\Big] \bigg\}.
\end{gathered}
\label{fkprim7}
\end{equation}
Here, we have used the formula
\begin{equation}
\displaystyle\int\limits_{0}^{t}e^{-\omega z} e^{-2A_1\tau_a e^{-z/\tau_a}}dz=\tau_a(2A_1 \tau_a)^{-\omega\tau_a} \cdot \Gamma\big[\omega \tau_a,2A_1\tau_a e^{-t/\tau_a},2A_1\tau_a\big], \qquad \omega\geq 0,
\label{fkprim8}
\end{equation}
where $\Gamma\big[\alpha,z_0,z_1\big]$ is the incomplete gamma function defined by
\begin{equation}
\Gamma\big[\alpha,z_0,z_1\big]=\displaystyle\int\limits_{z_0}^{z_1}z^{\alpha-1}e^{-z}dz.
\label{fkprim8a}
\end{equation}
Associated linear equation \eqref{lineq2}, \eqref{lineq2a}, \eqref{lineq2b} for this example reads
\begin{equation}
\begin{gathered}
\hat{{\mathcal{L}}}(\vec{x},t,{\bf C}_\varphi)v(\vec{x},t,{\bf C}_\varphi)=0,\\
v(\vec{x},t,{\bf C}_\varphi)\Big|_{t=0}=\varphi(\vec{x}),\\
L(t,{\bf C})=\tilde{a}(t)-\varkappa \sigma^2(t) \beta(t)\bigg(1-2\displaystyle\frac{D_{in}}{\varrho^2}-4\displaystyle\frac{D}{\varrho^2}\displaystyle\int\limits_{0}^{t}\tilde{D}_a(\tau)d\tau \bigg),\\
L_x(t,{\bf C})=0,\\
L_{xx}(t,{\bf C})=2\varkappa \sigma^2(t)\dac{\beta(t)}{\varrho^2}.
\end{gathered}
\label{fkprim9}
\end{equation}
Its solution can be obtained via the Green function of a parabolic equation as
\begin{equation}
\begin{gathered}
v(\vec{x},t,{\bf C}_\varphi)=e^{\frac{S(t,{\bf C}_\varphi)}{D}}\displaystyle\int\limits_{{\mathbb{R}}^n}G(\vec{x},\vec{y},t,{\bf C}_{\varphi})\varphi(\vec{y})d\vec{y},\\
S(t,{\bf C})=D\displaystyle\int\limits_{0}^{t}L(\theta,{\bf C})d\theta,\\
G(\vec{x},\vec{y},t,{\bf C})=\displaystyle\frac{A(t,{\bf C})}{\pi{\mathcal{D}}(t,{\bf C})}\exp\bigg[-\displaystyle\frac{(\vec{x}-\vec{y})^2}{{\mathcal{D}}(t,{\bf C})}-\dac{\vec{y}^2}{H(t,{\bf C})}\bigg],\\
A(t,{\bf C})=\exp\bigg[2\dil_{0}^{t}L_{xx}(\theta,{\bf C}){\mathcal{D}}(\theta)d\theta\bigg], \qquad H(t,{\bf C})=\bigg[\dil_{0}^{t}L_{xx}(\theta,{\bf C})d\theta\bigg]^{-1},
\end{gathered}
\label{fkprim10}
\end{equation}
where the function ${\mathcal{D}}(t,{\bf C})$ is the solution of the following Cauchy problem for the Riccati equation:
\begin{equation}
\dot{{\mathcal{D}}}=L_{xx}(t,{\bf C}){\mathcal{D}}^2+2D \tilde{D}_a(t), \qquad {\mathcal{D}}\big|_{t=0}=0.
\label{fkprim10a}
\end{equation}
The function ${\mathcal{D}}(t,{\bf C})$ in \eqref{fkprim10a} is a transcendental function. It can be seen that ${\mathcal{D}}(t,{\bf C})$ monotonously grows over time $t$ and ${\mathcal{D}}(t,{\bf C})={\rm O}(D)$.

Let us consider the Gaussian initial condition:
\begin{equation}
\varphi(\vec{x})=C_0\cdot\exp\Big[-\displaystyle\frac{\vec{x}\,^2}{D\gamma^2}\Big].
\label{fkprim11}
\end{equation}
Initial conditions for the Einstein--Ehrenfest system for \eqref{fkprim11} are as follows:
\begin{equation}
\sigma(0)=C_0 \pi D\gamma^2, \qquad D_{in}=\displaystyle\frac{D\gamma^2}{2}.
\label{fkprim12}
\end{equation}
Then, ${\bf C}={\bf C}_{\varphi}$, the integral \eqref{fkprim10} yields
\begin{equation}
\begin{gathered}
v(\vec{x},t)=\displaystyle\frac{C_0 A(t) H(t) D\gamma^2}{{\mathcal{D}}(t)\big(H(t)+D\gamma^2\big)+H(t)D\gamma^2}\exp\Big[\dac{1}{D}S(t)\Big]\times\\
\times \exp\Big[-\displaystyle\frac{\vec{x}\,^2}{{\mathcal{D}}(t)+\big(H(t)D\gamma^2\big)/\big(H(t)+D\gamma^2\big)}\Big],
\end{gathered}
\label{fkrim13}
\end{equation}
where the function $S(t)$ is given by \eqref{fkprim12}, \eqref{fkprim10}, \eqref{fkprim6}, \eqref{fkprim5a} and the argument ${\bf C}_{\varphi}$ is omitted for short.

Thus, the distribution of the ion/electron concentration is the diffusing Gaussin packet with the total quantity of ions/electrons determined by the \eqref{fkprim6}, \eqref{fkprim7}, \eqref{fkprim12}. Consider the qualitative behaviour of the solution \eqref{fkprim6}, \eqref{fkprim7} in details. The Einstein--Ehrenfest system \eqref{moment5a}, \eqref{moment5b}, \eqref{moment5c} is similar in the structure to the another one obtained for the Fisher--Kolmogorov-Petrovskii-Piskunov (FKPP) equation in \cite{fkppshap18}. The difference is that the equation for the zeroth-order moment $\sigma(t)$ had the quadratic nonlinearity in that work same as the FKPP equation opposed to the cubic nonlinearity in this work. It results in the qualitative difference of the solutions. For the FKPP equation, the zeroth-order moment can take negative values even for the positive initial condition that contradicts the physical meaning of the problem, so its interpretation is nontrivial. In this work, the zeroth-order moment $\sigma(t)$ \eqref{fkprim6}, \eqref{fkprim7} is positive over its entire domain. However, for some sets of the parameters, it exists only on a limited period of time. For physical reasons, the following conditions must be met for the equation \eqref{fkprim3}:
\begin{equation}
\bigg[1-4\displaystyle\frac{D_{in}}{\varrho^2}-8\displaystyle\frac{D}{\varrho^2}\displaystyle\int\limits_{0}^{t}\tilde{D}_a(\tau)d\tau\bigg]>0,
\label{fkprim14}
\end{equation}
since the triple recombination term would yield a negative contribution to the ion quantity in the active medium otherwise. Note that the condition \eqref{fkprim14} is violated at large times $t\sim\displaystyle\frac{1}{D}$ where $\sigma(t)\to 0$. The solutions \eqref{fkprim6}, \eqref{fkprim7} a priori exist at times where the condition \eqref{fkprim14} is met. Thus, the condition \eqref{fkprim14} is satisfied for any times if it is satisfied for $t\to \infty$. For \eqref{fkprim5a}, it yields
\begin{equation}
\bigg[1-4\displaystyle\frac{D_{in}}{\varrho^2}-8\displaystyle\frac{D}{\varrho^2}d_1\tau_d\bigg]>0,
\label{fkprim14a}
\end{equation}
that can be treated as an upper bound for the value of $(d_1 \tau_d)$ that meets the weak diffusion approximation. Otherwise, the asymptotic behaviour of the function $\sigma(t)$ at large times can be obtained by other method proposed in \cite{fkppshap18} assuming $u(\vec{x},t)=\tilde{u}(t)$. Since metal vapor active media are usually used in a pulse-periodic mode, the large times asymptotics are of little interest from the physical point of view and are not considered in this work.

Since the function \eqref{fkprim6}, \eqref{fkprim7} is given by the quite complex expression involving incomplete gamma functions, it has a number of behaviour types depending on parameters. We will focus on ones satisfying \eqref{fkprim14a}. The solution $\sigma(t)$ \eqref{fkprim6}, \eqref{fkprim7} can have two essentially different behaviour types. If the remanent temperature of electrons is sufficient for the excess ionization, then $\sigma(t)$ is the function with a single maximum point and the asymptote $\sigma=0$. In Fig. 1a, the plot of such the function is shown for $\varkappa=2$, $\varrho=0.5$, $\tau_a=\tau_b=\tau_d=1$, $A_1=1$, $d_1=2$, $B_2=0.4$, $B_1=0.2$, $D=0.01$, $D_{in}=0.01$, $\sigma(0)=1$. If the initial electron temperature is sufficiently small, then the function $\sigma(t)$ monotonically decrease tending to zero. This case is shown in Fig. 1b for same parameters except for $A_1=0.3$, $B_2=2$, $B_1=1$.

\begin{figure}[h]
  \begin{minipage}[b][][b]{0.45\linewidth}\centering
    \includegraphics[width=1\linewidth]{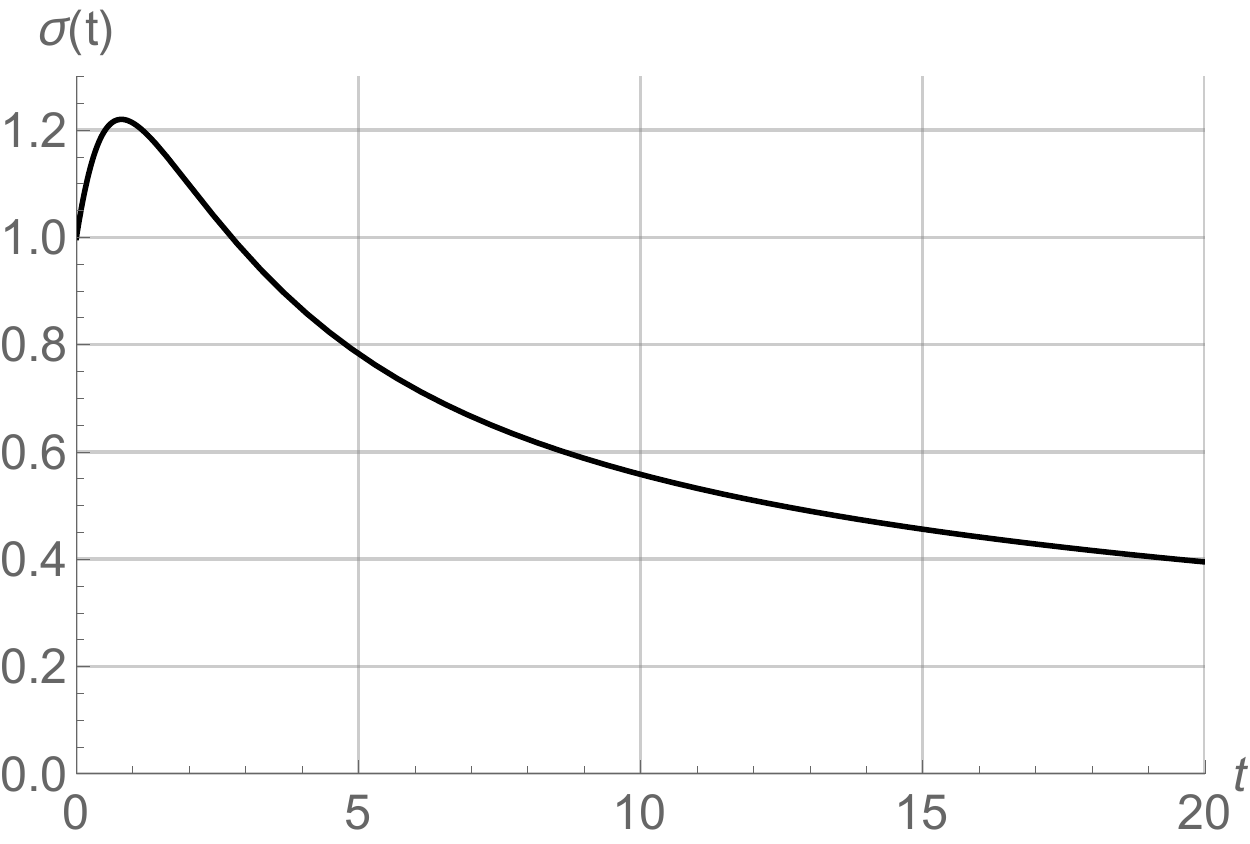} \\ a)
  \end{minipage}
  \hfill
  \begin{minipage}[b][][b]{0.45\linewidth}\centering
    \includegraphics[width=1\linewidth]{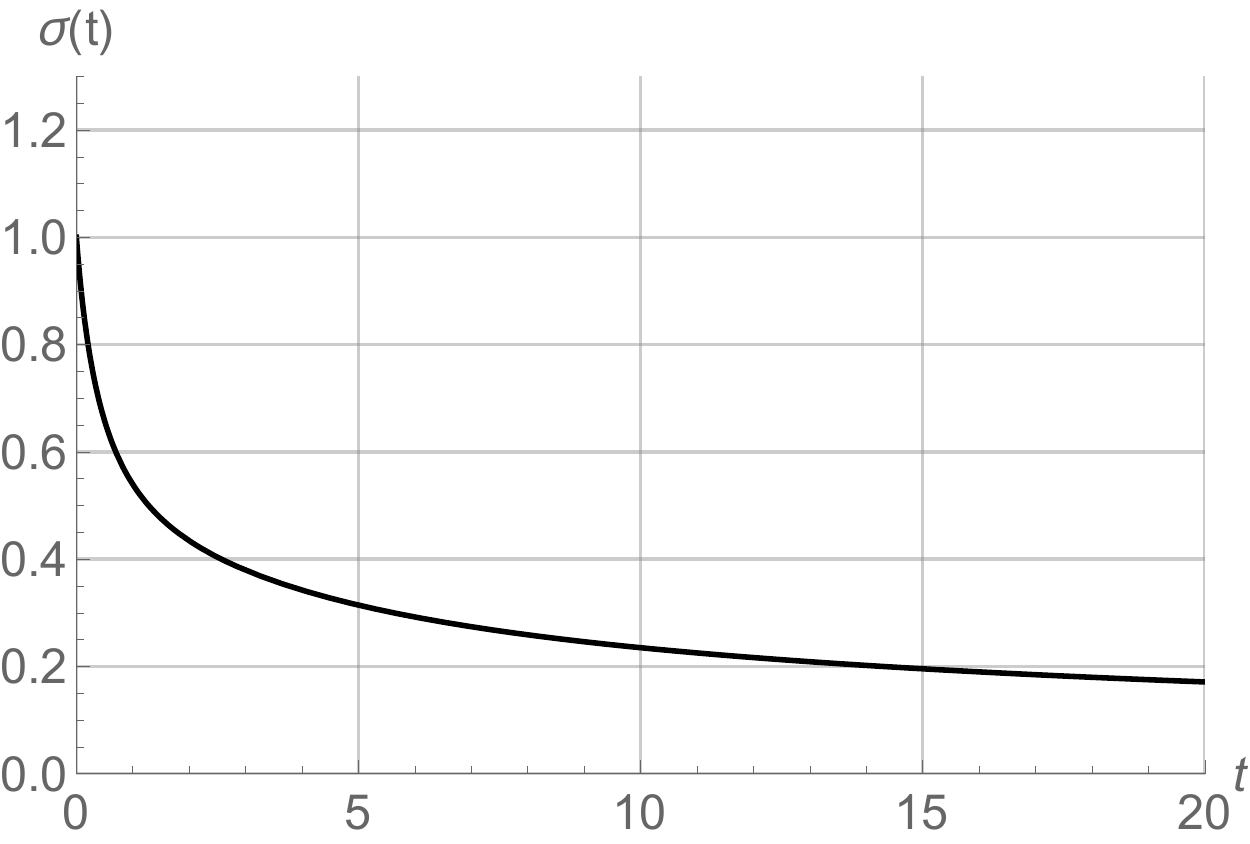} \\ b)
  \end{minipage}\\
  Fig. 1. The plot of the function $\sigma(t)$ for the high (a) and low (b) initial electron temperature.
\end{figure}

The bifurcation of behaviour types is determined by the sign of the number $I$ given by
\begin{equation}
I=A_1-\varkappa \sigma^2(0)B_1 \Big[1-4\dac{D_in}{\varrho^2} \Big]=A_1-\varkappa \sigma^2(0)B_1+{\rm O}(D).
\label{bif1}
\end{equation}
Thus, the condition $I>0$ ensures the presence of the excess ionization.

The formula \eqref{fkprim6}, \eqref{fkprim7} also admit one more kind of solutions satisfying \eqref{fkprim14a}, which is shown in Fig. 2 for the same parameters as in Fig. 1b except for $A_1=1.5$, $\tau_a=\tau_b=2$, $\tau_d=1$, $d_1=2.5$.

\begin{figure}[h]
\center{\includegraphics[width=0.45\linewidth]{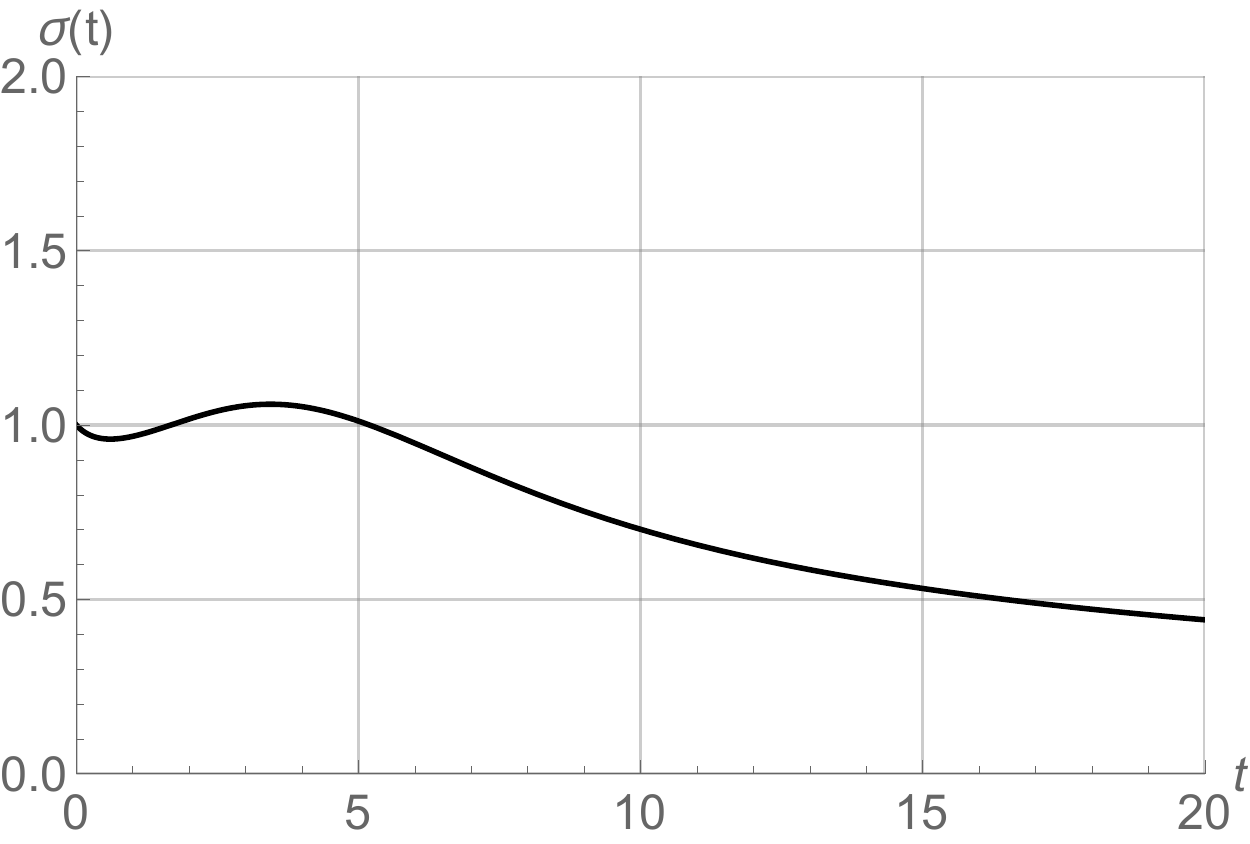}}\\
Fig. 2. The plot of the function $\sigma(t)$ for the special case.
\end{figure}

The solution in Fig. 2 have two extreme points. The prerequisite for such the case is that the condition
\begin{equation}
\left\{\begin{array}{l}
\dot{\sigma}(t)=0 \cr
\ddot{\sigma}(t)\geq 0
\end{array}\right.
\Longrightarrow
\dac{d}{dt}\frac{\beta(t)\bigg[1-4\displaystyle\frac{D_{in}}{\varrho^2}-8\displaystyle\frac{D}{\varrho^2}\displaystyle\int\limits_{0}^{t}\tilde{D}_a(\tau)d\tau\bigg]}{\tilde{a}(t)}\leq 0
\label{bif2}
\end{equation}
holds for some $t$. For $D\to 0$, the condition \eqref{bif2} yields
\begin{equation}
\dac{d}{dt}\dac{\beta(t)}{\tilde{a}(t)}\leq 0,
\label{bif3}
\end{equation}
that contradicts assumptions for the functions $\tilde{a}(t)$, $\beta(t)$ leading to \eqref{fkprim5a}. Therefore, the case shown in Fig. 2 is not of interest in the weak diffusion approximation and the sign of $I$ unequivocally determines the choice between cases shown in Fig. 1a and Fig 1b.

\end{section}

\begin{section}{Conclusion}
\label{sec:concl}

We have developed an approximate analytical approach based on the WKB--Maslov theory \cite{Maslov1}--\cite{BeD2} for studying kinetic phenomena in an active medium on metal vapors under the condition of quasi-neutrality in terms of the nonlocal kinetic equation  \eqref{fkpp2}.
The key point of the approach is the use of   the class  ${\mathcal P}_D^t$ of trajectory concentrated functions given by \eqref{fkpp4}
in which the solution to the Cauchy problem of equation \eqref{fkpp2} is sought.
This allows us to reduce the Cauchy problem for the kinetic equation \eqref{fkpp2} to the solution of the corresponding Cauchy problem
 for the associated linear equation \eqref{lineq1} or \eqref{lineq2}, using the general solution \eqref{gensol} of the EE system \eqref{moment7} of moments of the desired solution.
 As a result, we obtain  the leading term of the asymptotic solution of the Cauchy problem for the kinetic equation \eqref{fkpp2} accurate to ${\rm O}(D^{3/2})$ in the weak diffusion approximation.

Since the numerous  publications dedicated to the kinetic modeling of MVAM are mainly focused on numerical study, our approach can be used both for the approval of complex numerical models and as an independent method for calculation of the electron density in MVAM within given approximations.

The approach proposed here
can be considered as an extension of the method of semiclassical asymptotics in the class of functions ${\mathcal P}_D^t$, which we previously used in \cite{shap2009,fkppshap18} for the population nonlocal FKPP equation and for the nonlocal Gross--Pitaevsky equation in \cite{shapovalov:BTS1,sym2020}.

The solutions of the equation \eqref{fkpp3} obtained in this work have similarities with solutions of the FKPP equation constructed in \cite{shap2009,fkppshap18} for the one-dimensional case since these equations have the same set of stationary points for the spatially uniform functions $u\geq 0$: one unstable stationary point $u=0$ and one stable stationary point $u>0$.

 Nevertheless, the cubic nonlinearities leads to some distinctions. In particular, the zeroth-order moment $\sigma(t)$ in \eqref{fkprim6}, \eqref{fkprim7} for solutions \eqref{fkrim13} cannot take on a negative value for the positive initial condition while it could be for the asymptotic solutions of the FKPP  equation with the quadratic nonlinearity \cite{fkppshap18}. Since the zeroth-order moment corresponds to the population density in the FKPP model, it is not trivial how to interpret such the solutions from the physical point of view. The absence of this issue for the semiclassical approach to the model considered in this work means that this approach is more natural for the equation with the cubic nonlinearity. In this sense, the equation \eqref{fkpp3} is similar to the Gross--Pitaevskii equation \cite{shapovalov:BTS1,sym2020}.

 The plasma relaxation problem considered in Section \ref{sec:example} within the framework of the proposed method
 illustrates the construction of  the leading term of the semiclassical asymptotics for the kinetic equation \eqref{fkpp2} in explicit form using the incomplete gamma function. With the help of the  solution constructed, the time dependence of the number of ions $\sigma(t)$, which is an important characteristic of plasma kinetics, was obtained explicitly and analyzed. It is shown that the solution can correspond to the relaxation process with or without the excess ionization depending on the problem setup.

It can be seen from the results obtained that the WKB--Maslov method of semiclassical asymptotics can be
certainly
applied to the nonlocal generalization of the FitzHugh--Nagumo model \cite{fitz61,nagumo62} in the similar way as it was used for the two-component FKPP equation in \cite{fkppsym} and to the nonlocal generalization of the Zeldovich--Frank--Kamenetskii equation \cite{zeldovich85}. However, we expect the essentially different solution behaviour for them since those equations have different set of stationary points.

Since the asymptotic solutions of the nonlinear kinetic equation are found among the solutions of the associated linear equation, the nonlinear superposition principle for a family of such the solutions can be developed if the algebraic condition \eqref{alg} can be solved for them. The problem of constructing an infinite family of the analytical asymptotic solutions to the nonlinear kinetic equation \eqref{fkpp3} forming the basis for the nonlinear superposition principle is the future direction of our work. Also, we plan to generalize our approach to a two-component kinetic equation so that we can analyze more complex cases of a plasma behaviour.

\end{section}

\section*{Acknowledgement}
The reported study was funded by RFBR and Tomsk region according to the research project No. 19-41-700004. The work is supported by Tomsk State University under the International Competitiveness Improvement Program; by IAO SB RAS, project no. 121040200025-7.

\bibliography{lit1}

\providecommand{\newblock}{}
\begin{thebibliography}{10}
\expandafter\ifx\csname url\endcsname\relax
  \def\url#1{{\tt #1}}\fi
\expandafter\ifx\csname urlprefix\endcsname\relax\def\urlprefix{URL }\fi
\providecommand{\eprint}[2][]{\url{#2}}

\bibitem{kazaryan02}
Kazaryan M~A, Lyabin N~A and Zharikov V~M 2002 Technological systems based on
  copper vapor laser designed for measurement and material processing {\em
  Seventh International Symposium on Laser Metrology Applied to Science,
  Industry, and Everyday Life\/} vol 4900 ed Chugui Y~V, Bagayev S~N,
  Weckenmann A and Osanna P~H International Society for Optics and Photonics
  (SPIE) pp 1094--1098

\bibitem{asratyan16}
Asratyan A~A, Bulychev N~A, Feofanov I~N, Kazaryan M~A, Krasovskii V~I, Lyabin
  N~A, Pogosyan L~A, Sachkov V~I and Zakharyan R~A 2016 {\em Applied Physics A:
  Materials Science and Processing\/} {\bf 122} 434

\bibitem{klyuch19}
Klyuchareva S~V, Ponomarev I~V, Topchiy S~B, Pushkareva A~E and Andrusenko Y~N
  2019 {\em Journal of Lasers in Medical Sciences\/} {\bf 10} 350--354

\bibitem{lasmon14}
Evtushenko G~S, Trigub M~V, Gubarev F~A, Evtushenko T~G, Torgaev S~N and
  Shiyanov D~V 2014 {\em Review of Scientific Instruments\/} {\bf 85}

\bibitem{lasmon16}
Trigub M~V, Torgaev S~N, Evtushenko G~S, Troitskii V~O and Shiyanov D~V 2016
  {\em Technical Physics Letters\/} {\bf 42} 632--634

\bibitem{kulopt17}
Evtushenko G~S, Torgaev S~N, Trigub M~V, Shiyanov D~V, Evtushenko T~G and
  Kulagin A~E 2017 {\em Optics Communications\/} {\bf 383} 148--152

\bibitem{gubarev16}
Gubarev F~A, Trigub M~V, Klenovskii M~S, Li L and Evtushenko G~S 2016 {\em
  Applied Physics B: Lasers and Optics\/} {\bf 122}

\bibitem{behrouzina19}
Mohammadpour~Lima S, Behrouzinia S and Khorasani K 2019 {\em Applied Physics B:
  Lasers and Optics\/} {\bf 125}

\bibitem{boichenko05}
Boichenko A~M and Yakovlenko S~I 2005 {\em Laser Physics\/} {\bf 15} 1528--1535

\bibitem{marshall04}
Withford M~J, Brown D~J~W, Mildren R~P, Carman R~J, Marshall G~D and Piper J~A
  2004 {\em Progress in Quantum Electronics\/} {\bf 28} 165--196

\bibitem{kushner83}
Kushner M~J and Warner B~E 1983 {\em Journal of Applied Physics\/} {\bf 54}
  2970--2982

\bibitem{newell69}
Newell A~C and Whitehead J~A 1969 {\em Journal of Fluid Mechanics\/} {\bf 38}
  279--303

\bibitem{vaneeva19}
Vaneeva O, Boyko V, Zhalij A and Sophocleous C 2019 {\em Journal of
  Mathematical Analysis and Applications\/} {\bf 474} 264--275

\bibitem{Ji2008}
Ji A~C, Liu W, Song J and Zhou F 2008 {\em Physical Review Letters\/} {\bf 101}
  010402

\bibitem{physrev1}
Kasamatsu K, Tsubota M and Ueda M 2002 {\em Physical Review A - Atomic,
  Molecular, and Optical Physics\/} {\bf 66} 053606

\bibitem{Wang2010}
Wang D~S, Hu X~H, Hu J and Liu W 2010 {\em Physical Review A - Atomic,
  Molecular, and Optical Physics\/} {\bf 81} 025604

\bibitem{amara93}
Amara P, Hsu D and Straub J~E 1993 {\em Journal of Physical Chemistry\/} {\bf
  97} 6715--6721

\bibitem{Liang2005}
Liang Z, Zhang Z and Liu W 2005 {\em Physical Review Letters\/} {\bf 94} 050402

\bibitem{freund90}
Freund R~S, Wetzel R~C, Shul R~J and Hayes T~R 1990 {\em Physical Review A\/}
  {\bf 41} 3575--3595

\bibitem{gurpit64}
Gurevich A~V and Pitaevskii L~P 1964 {\em Soviet Physics JETP\/} {\bf 19}

\bibitem{carman98}
Carman R~J, Withford M~J, Brown D~J~W and Piper J~A 1998 {\em Optics
  Communications\/} {\bf 157} 99--104

\bibitem{boichenko2001}
Boichenko A~M, Evtushenko G~S, Yakovlenko S~I and Zhdaniev O~V 2001 {\em Laser
  Physics\/} {\bf 11} 580--588

\bibitem{kyureg19}
Kyuregyan A~S 2019 {\em Optics and Spectroscopy\/} {\bf 126} 388--393

\bibitem{yurchenko84}
Borovich B~L and Yurchenko N~I 1984 {\em Soviet journal of quantum
  electronics\/} {\bf 14} 1391--1400

\bibitem{carman94}
Carman R~J, Brown D~J~W and Piper J~A 1994 {\em IEEE Journal of Quantum
  Electronics\/} {\bf 30} 1876--1895

\bibitem{cheng97}
Cheng C and Sun W 1997 {\em Optics Communications\/} {\bf 144} 109--117

\bibitem{kulopt20}
Kulagin A~E, Torgaev S~N and Evtushenko G~S 2020 {\em Optics Communications\/}
  {\bf 460} 125136

\bibitem{kulopt19}
Torgaev S~N, Kulagin A~E, Evtushenko T~G and Evtushenko G~S 2019 {\em Optics
  Communications\/} {\bf 440} 146--149

\bibitem{kulphys18}
Kulagin A~E, Torgaev S~N, Evtushenko G~S and Trigub M~V 2018 {\em Russian
  Physics Journal\/} {\bf 60} 1987--1992

\bibitem{Maslov1}
Maslov V 1976 {\em Operational Methods\/} (Moscow: Mir Publishers)

\bibitem{Maslov2}
Maslov V 1994 {\em The Complex WKB Method for Nonlinear Equations. I. Linear
  Theory\/} (Basel: Birkhauser Verlag)

\bibitem{BeD2}
Belov V~V and Dobrokhotov S~Y 1992 {\em Theoretical and Mathematical Physics\/}
  {\bf 92} 843--868

\bibitem{shap2009}
Trifonov A~Y and Shapovalov A~V 2009 {\em Russian Physics Journal\/} {\bf 52}
  899--911

\bibitem{LST14}
Levchenko E~A, Shapovalov A~V and Trifonov A~Y 2014 {\em Journal of Physics A:
  Mathematical and Theoretical\/} {\bf 47} 025209

\bibitem{fkppshap18}
Shapovalov A~V and Trifonov A~Y 2018 {\em International Journal of Geometric
  Methods in Modern Physics\/} {\bf 15} 1850102

\bibitem{shapovalov:BTS1}
Belov V~V, Trifonov A~Y and Shapovalov A~V 2002 {\em International Journal of
  Mathematics and Mathematical Sciences\/} {\bf 32} 325--370

\bibitem{sym2020}
Shapovalov A~V, Kulagin A~E and Trifonov A~Y 2020 {\em Symmetry\/} {\bf 12} 201

\bibitem{kulagin2021}
Kulagin A~E, Shapovalov A~V and Trifonov A~Y 2021 {\em Symmetry\/} {\bf 13}
  1289

\bibitem{bagrov1}
Bagrov V~G, Belov V~V and Trifonov A~Y 1996 {\em Annals of Physics\/} {\bf 246}
  231--290

\bibitem{LST16}
Levchenko E~A, Shapovalov A~V and Trifonov A~Y 2016 {\em Journal of Physics A:
  Mathematical and Theoretical\/} {\bf 49} 305203

\bibitem{fitz61}
FitzHugh R 1961 {\em Biophysical journal\/} {\bf 1} 445--466

\bibitem{nagumo62}
Nagumo J, Arimoto S and Yoshizawa S 1962 {\em Proceedings of the IRE\/} {\bf
  50} 2061--2070

\bibitem{fkppsym}
Shapovalov A~V and Trifonov A~Y 2019 {\em Symmetry\/} {\bf 11} 366

\bibitem{zeldovich85}
Zeldovich Y~B, Barenblatt G~I, Librovich V~B and Makhviladze G~M 1985 {\em The
  Mathematical Theory Of Combustion And Explosions\/} (New York: Consultants
  Bureau)

\end{thebibliography}

\end{document}